# Metamorphic Malware Detection Using Linear Discriminant Analysis and Graph Similarity

Reza Mirzazadeh[1], Mohammad Hossein Moattar[2], Majid Vafaei Jahan[3]
Dept. of Computer Engineering
Mashhad Branch, Islamic Azad University,
Mashhad, Iran
[1]rmirzazadeh@mshdiau.ac.ir
[2]moattar@mshdiau.ac.ir
[3]vafaeiJahan@mshdiau.ac.ir

*Abstract*— The most common malware detection approaches which are based on signature matching and are not sufficient for metamorphic malware detection, since virus kits and metamorphic engines can produce variants with no resemblance to one another. Metamorphism provides an efficient way for eluding malware detection software kits. Code obfuscation methods like dead-code insertion are also widely used in metamorphic malware. In order to address the problem of detecting mutated generations, we propose a method based on Opcode Graph Similarity (OGS). OGS tries to detect metamorphic malware using the similarity of opcode graphs. In this method, all nodes and edges have a respective effect on classification, but in the proposed method, edges of graphs are pruned using Linear Discriminant Analysis (LDA). LDA is based on the concept of searching for a linear combination of predictors that best separates two or more classes. Most distinctive edges are identified with LDA and the rest of edges are removed. The metamorphic malware families considered here are NGVCK and metamorphic worms that we denote these worms as MWOR. The results show that our approach is capable of classifying metamorphosed instances with no or minimum false alarms. Also, our proposed method can detect NGVCK and MWOR with high accuracy rate.

I. INTRODUCTION

Today malware and viruses are serious problems for governments, organizations and individuals. Malware refers to software designed specifically to damage or disrupt a system [2]. A metamorphic malware is one that can transform based on the ability to translate, edit and rewrite its own code. Metamorphism is the process of transforming a piece of code into unique instances [1]. In metamorphic malware, copies of the instances are functionally equivalent, but their internal structures and source codes differ. This ability allows new variants to evade detection.

Signature scanning has been largely used as an antivirus technique. Current anti-viruses (AV) fail to detect metamorphic malware due to their varied internal structures. As malware writers are aware of the popularity of signature based AV, they have invented several techniques to evade signature-based detection [3]. These transformations include register renaming, code permutation, dead code insertion and block dead code insertion.

Metamorphic malware is considered more difficult to write than other malware such as polymorphic. In order to ease this difficulty, malware writers have developed virus creation kits. One of the most famous virus kits is "Next Generation Virus Creation Kit" (NGVCK) [5]. It can automatically generate new variants of a virus with the same behavior.

Many methods have been proposed to detect metamorphic viruses, which can be categorized into two families: those that use dynamic analysis and those that rely on static analysis of the code. [2]. Dynamic analysis refers to observing a malware's behavior during run-time while static analysis is the testing and evaluation of a malware by examining the code without executing it. Dynamic analysis is costly and needs an isolated environment to perform. Furthermore, it suffers from incomplete code coverage because it monitors only one execution path while static analysis covers all part of a file. In this study, we propose a method based on static analysis and similarity method.

In this paper, we investigate on the method proposed by Runwall et al. in [4]. In basic Opcode Graph Similarity (OGS) all nodes and edges contribute to the final result. Therefore, this approach is not immune against code obfuscation like dead code insertion. To address this problem, we combine the proposed method by Linear Discriminant Analysis (LDA) in order to prune dead codes from the graphs. Also, we used a more precise criterion to set a threshold. The results are promising and show high accuracy rate for detection of NGVCK and MWOR [6] metamorphic malware.

The paper is organized as follows: In section 2, related works are reviewed. In section 3 background information for the proposed method is provided. Then in the next section we present our methodology. In section 5, we illustrate our experimental result on NGVCK and MWOR. Section 6 discusses different aspects of proposed method. Finally section 7 contains our conclusion.

II. RELATED WORKS

Prior research in [7] developed a statistical method for metamorphic malware detection using Hidden Markov Model (HMM). The main idea was to train an HMM with opcodes extracted from viruses of a metamorphic family. The trained HMM will model the characteristics of a metamorphic virus. With this solution, it would be possible to calculate a score representing how close a file is to a virus family given by the trained HMM. Also, they revealed that NGVCK has the highest rate of obfuscation comparing with other metamorphism engines. In [8], authors proposed a method based on Profile HMM (PHMM) for metamorphic detection. A low detection rate was achieved for NGVCK but VCL-32 and PS-MPC detection rates were acceptable. Ref. [9] has studied on more obfuscated metamorphic malware and evaded HMM detector by inserting dead code of benign files to malware ones which resulted in poor accuracy. To tackle this problem, authors in [10] published a method based on statistical techniques which improved HMM in combination with Chi-squared test.

Many efforts have been made in accordance with malware detection by using graph analysis. Reference [11] proposes a graph-based method for malware detection. It extracts the sequence of opcodes and builds a weighted directed graph where each opcode is a node of the graph. Authors of [4] tried to improve the last method of metamorphic detection using Opcode Graph Similarity (OGS) and could obtain a high accuracy rate for NGVCK detection which was comparable to HMM detection rate.

Researchers have published a metamorphic worm in [12]. This malware is highly obfuscated and it can easily evade HMM and OGS. What distinguishes this metmorphic malware from the others is that it can carry its own morphing engine. Authors in [13] used LDA and data mining methods for metamorphic malware detection. Using LDA, they could rank opcode bi-gram features for classifying benign and malware files. The accuracy rate for NGVCK was about 99.7%.

In overall, prior researches are based on similarity methods and prone to elusion because they do not provide clear solution for encountering code obfuscation such as dead-code insertion.

### III. Materials and Methods

#### A. Linear Discriminant Analysis

Linear Discriminant Analysis (LDA) is a method used in many fields such as machine learning and pattern recognition for extracting features which preserve class separability. LDA is based upon the concept of searching for a linear combination of predictors that best separates two or more classes. Using LDA, features are selected based on the ratio of the total within-class variability and between-class variances. Within-class scatter matrix $S_w$ is computed by (1):

$$S_w = \sum_{i=1}^{C} P_k S_i \quad (1)$$

C denotes the number of classes. In our problem, it would be 2 because there are two classes, i.e. Malware and Benign. $P_k$ is the probability for class k and is considered 0.5 in this paper. $S_i$ is the variance of the features and computed as follows (2):

$$S_i = \sum_{x \in D_i}^{n} (x - m_i)(x - m_i)^T \quad (2)$$

Here $m_i$ is the mean vector, $x$ is the value of each feature and $S_i$ is the resulting scatter matrix of the ith class. Between-class variability matrix $S_B$ is computed by the (3) as follow:

$$S_B = \sum_{i=1}^{C} N_i (m_i - m)(m_i - m)^T \quad (3)$$

In (3) $m$ is the overall mean, and $m_i$ and $N_i$ are the sample mean and sizes of the classes, respectively. It can be clearly seen that between-class variability is computed by variance of class centers with respect to global center.

#### B. Opcode Graph Similarity

This method was introduced in [4]. It is based on a graph-based technique that was used in [14]. In this approach authors try to make a weighted directed graph from opcodes of binary files. Each distinct opcode is a node in a directed graph. For each transition between opcode nodes, a weighted directed edge is added. Edge weights are the transition probabilities. The dissimilarity score of A and B opcode graphs is computed as follows:

$$\text{Score}(A, B) = \frac{1}{N^2} \left( \sum_{i,j=0}^{N-1} |a_{ij} - b_{ij}| \right)^2 \quad (4)$$

Experiments indicate that malware graphs are more similar and they are different from benign graphs. Therefore, it is possible to set a threshold to distinguish malware and benign programs.

### IV. Proposed Methodology

Fig. 1 summarizes the proposed methodology. The main goal here is to improve the opcode graph similarity method with the aid of LDA. With regard to OGS technique, it is needed to make graphs from opcodes and compare the graphs in order to determine whether the input file belongs to the metamorphic family or not. In the following section, the proposed approach is described.

#### A. Preprocessing

In the first step we need to prepare our dataset. As mentioned before, our method is based on static analysis, so it is needed to extract opcode sequences of files. For extracting opcodes, each file should be converted to the machine instructions. We used IDA Pro [15] for this purpose. Operands in machine instructions do not have an important role, so we omitted them. Then, the dataset is divided into training and validation sets. In this study, k-cross-fold validation is used [16].

## B. Training

As mentioned before, in OGS, all edges and nodes play a respective part in the outcome. One of the most popular obfuscation techniques used in metamorphic malware is dead-code insertion; therefore, dead-code would be part of graphs and edges. It could lead to false alarms in OGS. According to [6], OGS is unable to distinguish high obfuscated metamorphic worms from benign programs. Hence, the main goal of this step is removing dead-code and junk edges from graphs.

In section 3, it was shown that LDA could calculate within and between variance of features. Our features here are edges and their weights. If we could find edges that have less within-class scatter and more between-class variability, then they would be good candidates to remain in the graph. Best edges are kept and the rest of them will be pruned.

In order to perform this process, all distinct edges and their weights should be extracted. Then, within and between scatterings of each edge are calculated. To find most effective edges, Eq (5) is used. We ranks edges in descending order based on the following $R_e$ value. Definitely low rank edges should be removed. So, with the aid of the threshold, most effective features are ready to be used.

$$R_e = \frac{S_W + S_B}{S_W} \qquad (5)$$

Now the distinctive edges are ready and the graphs will be pruned. In other words, all edges are removed from a graph except distinctive edges. With this technique, malware files would be more similar to each other because their garbage codes were removed and also differ from benign programs. Consequently, setting a discriminative threshold is more affordable.

## C. Set Treshold

The first and most important task in this phase is using distinctive edges in comparisons. A threshold can separate benign and metamorphic malware programs. In order to set a threshold, all pairs of training malware files should be compared with each other using Eq. (4) with regard to the distinctive edges. Afterwards, all training benign files are compared with all pairs of malware files. It is a strict and precise measure to confirm that all benign and malware files are sharply distinguished. In previous works such as [4] only adjacent files were compared. At the end of this step, there is a threshold that can distinguish malware and benign files. Now, the trained model is ready for use. In conclusion, in the trained model, malware files are similar and benign files are different from malware programs. Therefore, a threshold can differentiate them.

## D. Prediction

The last step of the proposed method is prediction. All newly entered benign and malware files in the test set are compared with the malware files in the trained model. It is expected that input malware be similar to the instances in trained model and stand under the threshold. Conversely, benign programs should stand above the threshold, and any threshold violation leads to false alarms.

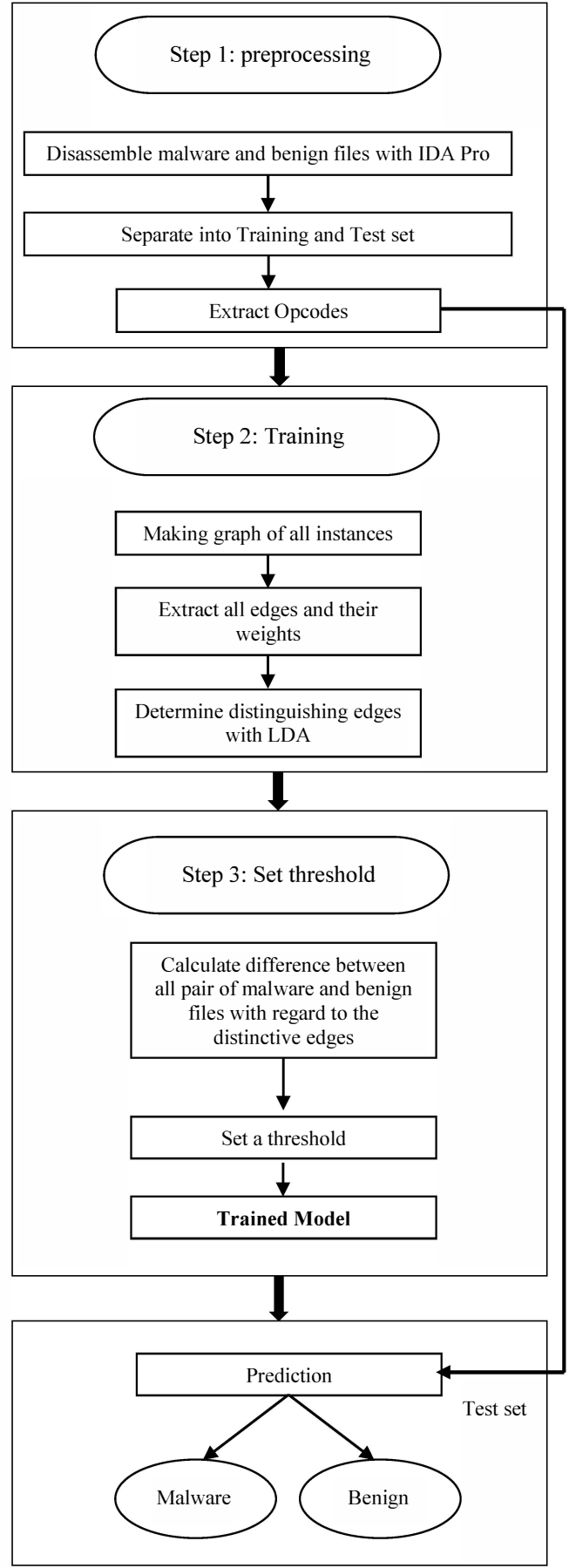

Figure1. Flow of proposed method

## V. EVALUATIONS

Improved OGS was implemented in Python and C programming languages. We conducted all the experiments on platform having 4 GB RAM and Core i5-M460@2.53 GHz processor and the operating system was 64-bit Windows 7. In the following section, we will discuss the experimental results.

### A. Dataset

The metamorphic malware families considered here are NGVCK [5] and worms developed in [6] and we denote these worms as MWOR. MWOR uses different methods for obfuscation. Dead-code insertion from benign files is widely used in MWOR. The authors of [6] define the ratio of dead-code to worm-code as the "padding-ratio". For example a padding ratio of 2 indicates that a worm has twice as much dead-code as worm instructions. Our dataset consists of 200 NGVCK malware and 40 benign files from Cygwin utility [17]. A wide variety of metamorphic detection approaches have used this dataset [7, 8, 10]. Furthermore, in [7], these malware files are shown to be the most highly metamorphic viruses generated with malware kits.

MWOR is a Linux-based malware; hence our benign files were selected form Linux operating system. It consists of 20 benign files. We used distinct set of MWOR files with padding ratios of 0.5, 1.0, 1.5, 2.0, 2.5, 3.0, 3.5, and 4.0. Totally there are 800 MWOR worms.

### B. Evaluation

There are 4 possible outcomes for detection. True Positive (TP) which is the number of infected files that are classified correctly, False Positive (FP) which is the number of benign files that are classified as malware. True Negative (TN) which is the number of benign files misclassified and False Negative (FN) which is the number of malware files that are classified as benign. The accuracy rate is the number of correct classification acquired divided by the total number of test files which is computed as follows:

$$Total\ Accuracy = \frac{TP + TN}{TP + TN + FP + FN} \quad (6)$$

In MWOR dataset there are 8 categories. In order to get a precise accuracy for each fold, (7) is used:

$$Mean\ Fold\ Accuracy = \frac{Total\ Accuracy + \frac{\sum TPR}{N}}{2} \quad (7)$$

We applied five-fold cross-validation in our experiments and used the mean of the accuracy values achieved from the folds, which is denoted as mean maximum accuracy (MMA) rate [10] and is computed as follows:

$$MMA = \frac{1}{5}\sum_{i=1}^{5} Accuracy_i \quad (8)$$

Where $Total\ Accuracy_i$ indicates the resulting accuracy of $i$th fold in cross-validation.

### C. Experimental results

In this section we present the result of our experiments using an improved OGS detector. As mentioned before, we used 5-cross-validation for NGVCK where the data is divided into five equal subsets. Each fold has 160 malware and 20 benign files for training and the rest of them are for the test. Fig. 2 shows the similarity score between malware-malware files and malware-benign files which are based on Eq (4) from a sample fold of training data. In this example the threshold is 8 and it can be clearly seen that both malware and benign files are classified correctly.

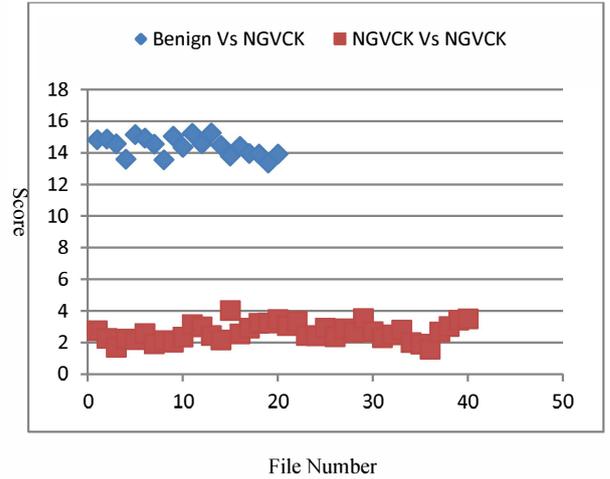

Figure 2. Similarity score for benign versus NGVCK

As mentioned, we compared our test data with all instances of training set. For example, for 40 malware samples in test set we conducted 6400 comparison between training set and test set and No threshold violation was found. Also, we conducted 3200 comparisons for benign files with metamorphic viruses in the training set and again the threshold was not violated. We repeated these steps for other folds and the same result achieved. Consequently, it can confirm that our proposed method is a strong approach for NGVCK detection. Table I illustrates the result of NGVCK detection for Improved OGS.

Table I  NGVCK detection results

| Fold | Accuracy (percent) |
|---|---|
| 1 | 100 |
| 2 | 100 |
| 3 | 100 |
| 4 | 100 |
| 5 | 100 |
| MMA | 100 |

For MWOR, we used two fold cross validation because there are 20 benign files and accurate result at least 10 files are needed for training phase. Therefore, 10 files from benign and 50 files from malware with padding ratio 0.5 were selected for training. The rest of the malware used for test. Fig. 3 shows the similarity scores of the malware and benign training samples. According to this result, the threshold is considered 4.

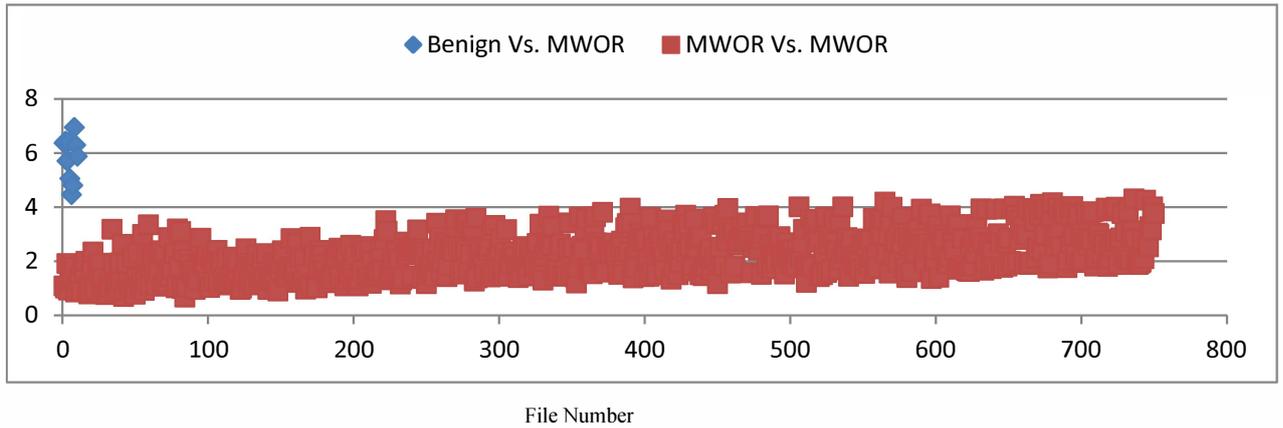

Figure 3. Similarity score for benign versus MWOR

Table II  MWOR detection results

| Fold | Mean Fold Accuracy (percent) | False Positive | False Negative |
|---|---|---|---|
| 1 | 99.07 | 0 | 0.01 |
| 2 | 99.79 | 0.01 | 0.001 |
| **MMA** | **99.43** | | |

In another experiment which is denoted in Table II, we compare test set data with random instances in training model. It can be clearly seen that with increasing obfuscation, the similarity of malware files be close to benign files. The total accuracy for this fold is about 99%.

*D. Discussion*

Fig. 4 illustrates the accuracy rate of NGVCK and MWOR detection for the proposed method and original OGS approach. This figure shows that the proposed enhancement on OGS approach (i.e. LDA opcode selection) was effective and the proposed approach is less vulnerable against detection of high metamorphic malware such as MWOR.

In our experiments we showed that our method can effectively detect NGVCK and MWOR. There are some parameters which can be discussed. During the experiments, we maintained 50 most discriminative edges. In our complementary experiments on NGVCK we changed the number selected top edges. Table III shows the results of these experiments for NGVCK. The results show that FP and FN happens when 200 edges are maintained.

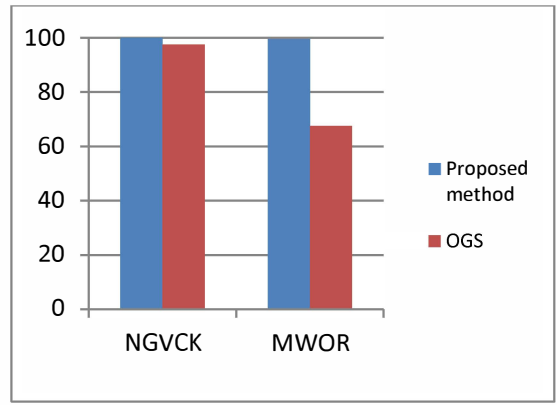

Figure 4. Accuracy rate of the proposed method versus OGS for NGVSK viruses and MWOR worms

Table III. No. of selected top edges vs. accuracy rate

| Top Edges | Accuracy rate |
|---|---|
| 50 | 100 |
| 100 | 100 |
| 150 | 100 |
| 200 | **99** |

Table IV shows our approach and well-known methods and their results for metamorphic malware detection. Structural Entropy [21] is a strong method for MWOR detection but according the experimental result, it cannot detect NGVCK successfully. This approach applies directly to binary files and structural entropy score depends on segment length and the number of segments. NGVCK tends to produce a lot of segments. Therefore, it is successful for NGVCK detection. In [22] Simple Substitution Distance (SSD) is proposed based on substitution cipher cryptanalysis. The result is not very promising for MWOR with high padding ratio although area under the curve (AUC) is quite well for lower padding ratio.

Table IV  Comparison between well-known methods for NGVCK and MWOR detection and the proposed approach

| Method | NGVCK | MWOR |
|---|---|---|
| Proposed Method | ✓ | ✓ |
| OGS | ✓ | ✗ |
| Structural Entropy | ✗ | ✓ |
| HMM | ✓ | ✗ |
| SSD | ✓ | ✗ |

In [19] authors illustrate that OGS is a reliable method for Javascript metamorphic malware detection. It has better accuracy than HMM and other well-known methods for metamorphic malware detection. In [20], a real world application has been designed for Javascript metamorphic detection based on OGS. In this paper, we proved that our method has better performance than OGS. It is plausible that our method is a practical solution for web-based metamorphic malware detection.

I. CONCLUSION

Metamorphic malware detection is a very challenging research area, which has gained much attention during previous years. In this study, we proposed a similarity method based on Opcode Graph Similarity and Linear Discriminant Analysis for metamorphic malware detection. Our approach overcomes weakness of OGS [4]. In state-of-the-art OGS, all nodes and edges contribute to the final result but in the proposed improved OGS, with the aid of LDA, junk edges are pruned from graphs. In other words, dead-code opcodes will be removed. Therefore, benign-benign and malware-malware similarity will increase and setting a threshold would be possible. The proposed method yielded 100% total accuracy for NGVCK and 99% accuracy for MWOR malware and proves that Improved OGS is highly efficient for metamorphic malware detection.

Future work could include detailed examination on the parameters of the proposed approach. In addition, with regard to [19], it is possible that Improved OGS would have better accuracy for Javascript malware detection. Therefore, using our method for Javascript malware detection could leads to interesting result.

REFERENCES

[1] M. Stamp, Information Security: Principles and Practice. Wiley, New York (2011)

[2] J. Aycock, Computer Viruses and Malware. Springer Science+Business Media, LLC, 2006.

[3] P. Szor, The Art of Computer Virus Research and Defense, 1 edn. Addison Wesley Professional, Boston (2005).

[4] N. Runwal, R. Low, M., M. Stamp, "Op-code graph similarity and metamorphic detection," J. Comput. Virol. 8: 37–52 (2012).

[5] Heavens, V.X.:http://vx.netlux.org/.

[6] S.M. Sridhara, M. Stamp, "Metamorphic worm that carries its own morphing engine," J. Comput. Virol. Hacking Tech. 9(2), 49–58 (2013).

[7] W. Wong and M. Stamp, "Hunting for metamorphic engines," Journal in Computer Virology, vol. 2, no. 3, pp. 211-229, 2006.

[8] S. Attaluri, S. McGhee, M. Stamp,"Profile hidden Markov models and metamorphic virus detection," J. Comput. Virol. 5(2), 151–169 (2009).

[9] D. Lin, M. Stamp, "Hunting for undetectable metamorphic viruses," J. Comput. Virol. 7(3), 201–2014 (2011).

[10] A.H. Toderici, M. Stamp, "Chi-squared distance and metamorphic virus detection," J. Comput. Virol.9, 1–14 (2013)

[11] B. Anderson, et al. "Graph-based malware detection using dynamic analysis," J. Comput. Virol.7(4), 247–258 (2011)

[12] S.M. Sridhara, M. Stamp, "Metamorphic worm that carries its own morphing engine," J. Comput. Virol. Hacking Tech.9(2), 49–58 (2013)

[13] J. Kuriakose, P. Vinod, "Ranked Linear Discriminant Analysis Features for Metamorphic Malware Detection," In Proceedings of 4th IEEE International Advanced Computing Conference (IACC-2014

[14] B. Anderson, et al. "Graph-based malware detection using dynamic analysis," J. Comput. Virol.7(4), 247–258 (2011).

[15] IDA Pro. http://www.hex-rays.com/idapro/

[16] Geisser, S.: Predictive Inference: An Introduction. Chapman and Hall, London (1993)

[17] Cygwin. http://www.cygwin.com

[18] B. Rad, M. Masrom, and S. Ibrahim. "Opcodes Histogram for Classifying Metamorphic Portable Executables Malware," In ICEEE, pages 209--213, September 2012.

[19] M. Musale, T. Austin, M. Stamp, "Hunting for metamorphic JavaScript malware," (2014)

[20] S.R. Javaji, "firefox add-on for metamorphic javascript malware detection," (2015). Master's Projects. Paper 401.

[21] D. Baysa, et al."Structural entropy and metamorphic malware," J. Comput. Virol. 9(4), 179-192 (2013)

[22] G. Shanmugam, R.M. Low, M.Stamp, "Simple substitution distance and metamorphic detection," J. Comput. Virol. Hacking Tech. 9(3), 159–170 (2013)